\documentclass{article}

\usepackage{PRIMEarxiv}

\usepackage[utf8]{inputenc}
\usepackage[T1]{fontenc}
\usepackage[numbers,sort&compress]{natbib}
\usepackage{hyperref}
\usepackage{url}
\usepackage{booktabs}
\usepackage{amsfonts}
\usepackage{amsmath}
\usepackage{nicefrac}
\usepackage{microtype}
\usepackage{fancyhdr}
\usepackage{graphicx}
\graphicspath{{media/}}

\usepackage{titlesec}
\titlespacing*{\section}{0pt}{1.2ex}{0.8ex}

\title{Coding Agents Aren't Enough! \\ Evaluating an Enterprise Security Brain for Agentic Cloud Investigations}

\author{
  \textbf{Leon Goldberg},
  \textbf{Gal Engelberg},
  \textbf{Eden Yavin},
  \textbf{Elad Elouz},
  \textbf{Ariel Zadok},
  \textbf{Konstantin Koutsyi}\\
  Sola Security, Tel Aviv, Israel \\
  \texttt{Corresponding author: gal.e@sola.security}
}

\begin{document}
\maketitle

\begin{abstract}
Cloud-security investigation is dominated by \emph{population} tasks: which identities can read a data store, how many resources fail a control, which assets are reachable from another account. These resolve against a complete inventory, not a named object. A partial answer to one is not a partial result. It is a different result. General-purpose coding agents can now be given read-only cloud credentials and asked to investigate directly, which raises the question of what a purpose-built security context layer still contributes. We evaluate the Sola Security Brain, a security intelligence layer whose relational substrate is resolved offline and whose security logic is evaluated against it at query time, against two coding agents operating the same live AWS environment through a read-only CLI, Claude Code and OpenAI Codex, over 28 cloud-security investigation tasks. Answers are scored by a blinded, tier-weighted, grounding-gated \emph{relative recall} over a single joint claim pool built from all three systems, so the three scores share one denominator. The Sola Security Brain reaches $0.549 \pm 0.012$ coverage against $0.340 \pm 0.011$ for Claude Code and $0.281 \pm 0.006$ for Codex, across three grading draws whose ordering is identical in every draw. It leads 24 of 28 tasks from the cheapest model tier, at $17.8\times$ and $20.8\times$ lower cost per task than Claude Code and Codex, and $28.7\times$ and $40.6\times$ lower cost per unit of coverage. Beyond the aggregate, we describe an answer-level pattern we term \emph{sample-and-generalise}: under a turn budget both coding agents enumerate a fraction of a large population, assert an unhedged universal negative, and disclose the sample size only in answer metadata rather than in the answer. In one task Claude Code reported that no bucket policies exist after checking four bucket families, in a sweep that sampled 40 of roughly 5{,}000 buckets, in an account where 65 buckets carry a wildcard-principal read grant.
\end{abstract}


\section{Introduction}\label{sec:intro}

A security analyst asking ``which roles can read my storage buckets'' is not asking for a lookup. The answer is a property of a population: every identity-based grant, every resource policy, every wildcard or administrative entitlement that subsumes the specific one, resolved across an estate that may contain thousands of buckets and tens of thousands of roles. The same is true of most of the working questions in cloud security. \emph{What should I fix first. Which identities give the widest reach if compromised. Which resources are reachable from another account.} Each is an aggregate over a complete set.

Two developments have changed how these questions get answered. First, general-purpose coding agents have become competent operators of cloud APIs: given read-only credentials and a command-line interface, such an agent will plan an investigation, issue calls, follow what it finds, and write up a result. Second, purpose-built security platforms increasingly maintain a \emph{context layer} in which the substrate those relationships depend on, a normalised inventory with entities reconciled and relations validated, is resolved offline into a queryable representation, so that effective access, reachability, blast radius, failing controls and cross-account trust can be evaluated against it when a question arrives. The two approaches answer the same question from different positions: one reconstructs the relevant structure at question time from raw APIs, the other queries a structure that already exists.

That structured context improves LLM performance on security tasks is established, including by our own prior work. The Cross-Vendor Sola ISPM Benchmark held models and tasks fixed and varied only the injected context, and found that structured relational context raised answer correctness by roughly 34\% relative while cutting exploration queries by roughly 70\% \cite{Yavin2026_CrossVendorISPM}. Independent work reports the same direction in other domains. Supplying runtime topology to a patch-generating model raises correctness on topology-dependent Kubernetes findings from 11.1\% to 78.0\%, with no movement on a topology-independent control arm \cite{Shaikh2026_KuTIE}. Replacing a flat document store with a typed knowledge graph, holding the model constant, moves an industrial asset-operations agent from 65\% to 82--83\% \cite{Mandarapu2026_KGAssetOps}. We therefore treat the context effect as established and ask a different question of it.

The open question is \emph{where} that effect comes from on open-ended investigation, and what a system gains from the context it has. Existing agentic security benchmarks put the exploring agent under test rather than against a context layer \cite{Wu2026_ExCyTInBench,Wang2026_SecRespond,Chona2026_CyberDefenseBenchmark}. Coverage deficits in broad information-seeking are documented in general settings \cite{Wong2025_WideSearch}, and premature stopping is a named failure mode in multi-agent systems \cite{Cemri2025_MAST}. Neither literature examines what an agent \emph{asserts} when its budget runs out with a population only partly enumerated.

In this work we ask how far a context layer closes that gap. We evaluate the Sola Security Brain, in which a security intelligence layer is built offline and evaluated at query time, against two coding agents operating the same live AWS environment through a read-only CLI: Claude Code and OpenAI Codex. All three answer the same 28 investigation tasks, authored from that environment and validated by a security practitioner. We compare the three on \emph{coverage}: the share of the findings a complete answer would contain that each system actually surfaced.

We score coverage because these tasks are aggregates. What an analyst needs from the answer is how many of the identities holding read access it named, how many failing resources it listed, how many exposed paths it traced. An answer that is correct about the fifth of the population it examined is still wrong about the estate, and a correctness-style measure would score it well; coverage does not.

We measure coverage with pooled relative recall \cite{Clarke1997_RelativeRecall}, the standard construction for scoring systems against a merged judged set. Each answer is reduced to the atomic factual claims it makes about the environment, and the three systems' claims for a task are merged into one pool, relabelled and shuffled so the judge cannot tell which system produced which. Every claim in the pool is then rated for security consequence and weighted accordingly, so that the score tracks the share of security value surfaced rather than the number of statements made. Claims are validated before they count towards a system's score. Coverage is then a system's tier-weighted, validated share of the pool, and because the pipeline is itself model-driven, every figure we report is the mean of several independent grading draws. Section~\ref{sec:protocol} gives the measure in full and states its limits.

The rest of the paper is organised as follows. Section~\ref{sec:related} situates the work against agentic security benchmarks and documented agent coverage and budget failures. Section~\ref{sec:setup} describes the environment, the task set and the architectures under comparison. Section~\ref{sec:protocol} specifies the scoring protocol, its provenance in the measurement literature, and the efficiency measures. Section~\ref{sec:results} reports coverage, duration and cost, with per-theme and per-task decomposition. Section~\ref{sec:discussion} interprets the result, describes the answer-level pattern we observed behind it, and states the study's limitations. Section~\ref{sec:conclusion} concludes.

\section{Related Work}\label{sec:related}

Two areas bear directly on this work: benchmarks that place LLM agents in security investigation settings, and documented coverage and budget failures of tool-using agents. The measurement literature our protocol draws on is discussed with the protocol, in Section~\ref{sec:protocol}.

ExCyTIn-Bench is the closest structural analogue to our coding-agent arms: investigation questions derived from graphs over 57 log tables of a controlled Azure tenant, with the agent discovering the schema and investigating through SQL for a best reward of 0.606 \cite{Wu2026_ExCyTInBench}. Its graph authors and scores the questions but is never given to the agent, leaving the graph-as-context condition untested there. Others vary the setting: a forensic disk image with product alerts \cite{Wang2026_SecRespond}, an unhinted event database \cite{Chona2026_CyberDefenseBenchmark}, triage separated from novel-finding discovery \cite{Begimher2026_SIRBench}, a replayed cluster snapshot \cite{Wang2026_CloudOpsBench}, host-telemetry incident response \cite{Anand2026_AuditBench}, attack-chain reconstruction scored per reference step, where the strongest configuration recovers 39.6\% \cite{Liu2026_DiagChain}, and threat hunting decomposed into a dependency graph of analytical tasks \cite{Meng2026_CyberTeam}. All report low absolute performance on open-ended discovery, all put the exploring agent under test rather than against a context layer, and all score against a reference label set, not a pooled finding set.

Two results bear on our design, not our framing. ITBench instantiates incidents in live Kubernetes clusters, and its compliance and security track is the closest prior case of an agent given live infrastructure for posture work; agents resolve 25.2\% of those scenarios \cite{Jha2025_ITBench}. Cost-aware evaluation finds defensive investigation scales poorly with budget: doubling an agent's spend cap moved its defensive score by less than a point while tool calls rose by a tenth \cite{Kassianik2026_CostAware}. That is our boundary from another direction, the constraint being what an agent can reach, not how long it thinks. Cloud posture and effective-permissions analysis is otherwise thin, and the substantial coverage-oriented treatment of cloud IAM is not an agent at all: whitebox and greybox analysis over permission-flow templates, benchmarked on whether known privilege-escalation paths are found \cite{Hu2026_TACBench}. Our own prior work is the immediate context here, spanning identity-posture visibility on a live multi-vendor estate \cite{Engelberg2026_SolaVisibility}, a cross-vendor benchmark varying injected context with models and tasks fixed \cite{Yavin2026_CrossVendorISPM}, and a shared environment exposing enterprise security state through both a relational snapshot and live vendor APIs \cite{Engelberg2026_OSB}. This paper differs in what it compares: not an agent against itself under varying context, but a context layer against general coding agents on the same environment.

On the second area, coverage failures are documented. More than ten agentic systems score near zero on exhaustive collection over a large item set where humans approach completeness \cite{Wong2025_WideSearch}; premature termination and unawareness of stopping conditions are named failure modes \cite{Cemri2025_MAST}; raising a tool-call cap alone does not raise accuracy, since the agent cannot see how much allowance remains \cite{Liu2025_BudgetAwareToolUse}; and more searching makes abstention on unanswerable queries worse \cite{Xie2026_OverSearching}. Where the answer is a complete set, recall-oriented suites show the gap directly \cite{Rafiee2026_TotalRecallQA}, and exhaustive-list benchmarks name both premature stopping and the hedging that inflates recall with low-confidence items \cite{Gupta2026_DeepSearchQA}.

Nearer our observation is work on what an agent \emph{asserts} when its evidence does not cover the question. Frontier agents asked to review a file set fail to read all of it in 67.9\% of runs and, in 80.4\% of those, either claim a complete review or leave the gap undisclosed, missing planted defects at roughly 1.8 times the rate of agents that read everything \cite{Smyth2026_Overclaiming}. The epistemic error is named and measured in static settings as \emph{over-closure}, answering a universal query negatively on evidence that does not cover its scope \cite{Min2026_WhenAbsenceIsEvidence}. False absence claims are a category in hallucination detection for tool-using agents \cite{Basu2026_FalseAbsence}, search agents terminate with constraints unverified \cite{Ko2026_IllusoryCompletion}, agents read the lack of an immediate match as proof of non-existence \cite{Tang2026_SciExplore}, and frontier models reason poorly about when to decline \cite{Kirichenko2025_AbstentionBench}.

Our contribution on this point is specific to security. None of that work is set in a domain where the population is an organisation's live resource inventory and the universal negative states that a class of exposure does not exist, and the asymmetry matters: a file missed in code review is caught on the next pass, whereas an exposure reported as absent is not revisited. Nor does any of it examine where the disclosure goes. Overclaiming is measured as whether the gap was disclosed at all; we observe agents that do disclose the sample size, into answer metadata, not into the sentence carrying the claim.

\section{Experimental Setup}\label{sec:setup}

This section describes the three systems under comparison and the conditions they were run under. Section~\ref{subsec:brain} describes the enterprise security brain and what its representation makes available at inference time, Section~\ref{subsec:agent} the coding-agent baselines, and Section~\ref{subsec:env} the environment, the task set and the run configuration. No architecture is described here as a systems specification; nothing in the evaluation depends on implementation detail.

\subsection{The enterprise security brain}\label{subsec:brain}

A security brain is a context layer over an organisation's security-relevant state. It is the representation findings are derived from, built so that any question about the environment can be answered from it, including questions no policy anticipated. The conceptual architecture is set out in \cite{Engelberg2026_AINativeAssetIntelligence,Goldberg2026_ContextIsArchitecture}; we summarise the parts that bear on this evaluation. It has two cooperating layers, one that builds the representation and one that scores it.

The \emph{modelling} layer builds the representation. Provider-specific extractors normalise heterogeneous API responses into a vendor-agnostic tabular form held in an immutable, versioned store, so that any downstream computation is reproducible against an explicit snapshot. Those records are then lifted into a typed property graph. Its nodes are assets: identities, workloads, data stores, network endpoints, code artefacts. Its edges are the relations between them, including ownership, identity assumption, data flow, network reachability, configuration attachment and control attestation. Edges arise three ways: deterministic schema-driven rules encoding known intra-vendor semantics, joins on normalised identifiers that bridge otherwise disconnected vendor silos, and model-assisted synthesis in which candidate relations are proposed from the joint schema and then validated against actual data before acceptance. Assets from different vendors are connected through typed relations rather than isolated in per-vendor silos.

The graph carries more than inventory and topology. Findings from scanners and third-party security products are attached to the assets they concern, so that a finding is a property of a modelled asset rather than a row in a separate queue, and its vendor-assigned severity can be refined against the context the graph supplies. Semantic attributes describe what an asset is to the organisation rather than what it is technically: its business function, the criticality of the data it holds, the environment it belongs to. Configuration anomaly scores, measuring how far an asset's configuration departs from that of comparable assets, attach to the same assets. Above all of this sits derived security knowledge, the control predicates, attack-vector traversals and blast-radius patterns, held as executable specifications rather than materialised results, and evaluated against the underlying data at inference time.

The \emph{scoring} layer turns that representation into calibrated signals, and is deliberately conservative about where model judgement enters. Intrinsic exposure is measured from observed weaknesses and exploitability patterns, while contextual importance describes what the asset is to the organisation. The two combine multiplicatively. The model is not permitted to assign a final score; it adjusts intermediate semantic variables that a deterministic aggregation then consumes.

What is resolved before a question arrives is therefore the substrate rather than the answer. An agent querying this layer still investigates: it plans, queries, follows what it finds, and refines. What it does not have to do is discover a schema, reconcile the same principal appearing under four different identifiers in four systems, infer that one entitlement subsumes another, or invent a traversal for reachability. That work is done once, not per question.

\subsection{The coding-agent architecture}\label{subsec:agent}

Several general-purpose coding agents can be pointed at live infrastructure in the way these arms require. Command-line agents such as Claude Code \cite{Anthropic_ClaudeCode}, Codex CLI \cite{OpenAI_CodexCLI} and Gemini CLI \cite{Google_GeminiCLI} run in a terminal with shell access. Editor-integrated agents such as Cursor \cite{Anysphere_Cursor} and GitHub Copilot's agent mode \cite{GitHub_CopilotAgentMode}, and open-source agents such as Aider \cite{Gauthier_Aider} and OpenHands \cite{Wang2025_OpenHands}, expose the same underlying loop of plan, act, observe and revise. They differ in interface, in the model behind them and in how tool use is mediated, but not in the architectural position they occupy: none maintains a model of the environment between questions, and each rebuilds whatever structure a task needs from the tools available to it at the time. We evaluate the two most widely used, Claude Code \cite{Anthropic_ClaudeCode} and OpenAI Codex \cite{OpenAI_CodexCLI}.

The tested configuration is each agent given read-only credentials for the environment and the cloud provider's command-line interface as its tool. Claude Code runs on Claude Opus 4.8 and Codex on GPT-6-Astra. The two share a harness: both Codex and Claude Code given the same system prompt verbatim, with only a credentials-phrasing fix. Each receives the task text and nothing else: no schema, no inventory, no indication of where to look. From there each reconstructs whatever structure the question needs at the moment it is asked, discovering what exists by listing it, deciding which objects are worth inspecting, issuing the calls, and assembling the relationships itself, all within a finite budget of turns and time.

Two properties follow from that position. The agent reads current state directly rather than a representation resolved earlier, so what it returns reflects the environment as it stands at the moment of the call. And it is not restricted to relations someone anticipated modelling: any relationship derivable from the API surface is available to it, whether or not a schema was designed to hold it.

The budget is the binding constraint. Where a question ranges over a large population, the available turns and time bound how many members of that population can be inspected, so the agent selects a subset and decides where to spend its calls.

\subsection{Environment and task set}\label{subsec:env}

The evaluation environment is a real-life enterprise AWS environment: a live, multi-account estate of the kind an analyst works in daily, not a synthetic or seeded lab. It contains no production customer data. All three systems were given read-only access to it and answered the same tasks against it.

We elicited the tasks in two stages. A task-authoring agent, separate from every system under evaluation, surveyed the environment's data and proposed candidate tasks grounded in what was actually present; a human security practitioner then reviewed each candidate for relevancy, discarding tasks that were artificial, redundant or unrepresentative of analyst work. The task set was fixed before any run and left unchanged throughout. All three received identical task text, with no hints as to schema, expected answer shape, or where to look. The phrasing is deliberately that of an analyst, not a benchmark: short, natural, and free of completeness-signal words that would cue any system to enumerate. Table~\ref{tab:tasks} lists all 28 tasks across six themes.

\begin{table}[ht]
\centering
\caption{The 28 investigation tasks.}
\label{tab:tasks}
\small
\begin{tabular}{p{2.6cm} r p{9.6cm}}
\toprule
Theme & $n$ & Tasks \\
\midrule
Risk triage & 7 &
q01 What are my riskiest resources right now, and why? \newline
q02 Who are my top 10 riskiest IAM identities? \newline
q03 What are my riskiest compute assets? \newline
q04 What are my riskiest storage assets? \newline
q05 What should I fix first? \newline
q06 What are my highest-risk resource groups, and why? \newline
q07 What are my riskiest secrets and encryption keys? \\
\addlinespace
Secrets and keys & 4 &
q08 Which secrets can more than one identity read? \newline
q09 Rank my secrets by how many principals can read them. \newline
q10 If a workload is compromised, which secrets does it expose? \newline
q11 Which internet-facing workloads can reach my secrets? \\
\addlinespace
Access and reachability & 5 &
q12 Which IAM roles can read my S3 buckets? \newline
q13 Who can decrypt my KMS keys? \newline
q14 Which principals have wildcard access, and what can they reach? \newline
q15 Which Lambda functions can be invoked, and by whom? \newline
q16 For my riskiest role, what can it reach? \\
\addlinespace
Blast radius & 4 &
q17 Rank my resources by blast radius. \newline
q18 If my most exposed resource is compromised, how far does it reach? \newline
q19 Walk me through my riskiest cluster. \newline
q20 Which identities give the widest reach if compromised? \\
\addlinespace
Control failures & 4 &
q21 Which resources fail public-access controls, broken down by control? \newline
q22 Rank my resources by number of failing controls. \newline
q23 Which of my S3 buckets are publicly accessible? \newline
q24 Which resources fail encryption-at-rest? \\
\addlinespace
Exposure and cross-account & 4 &
q25 What are my most internet-exposed assets? \newline
q26 Which exposed resources can reach sensitive data? \newline
q27 What are the attack paths to my privileged roles? \newline
q28 Which resources are reachable from another account? \\
\bottomrule
\end{tabular}
\end{table}

The six themes cover the working surface of cloud-security analysis.

\begin{itemize}
\item \textbf{Risk triage.} Ranking problems: riskiest resources, identities, compute, storage, resource groups, secrets. Producing an ordering means scoring every candidate on the same basis, so a system that inspects a subset does not return a worse ranking. It returns a ranking of the wrong set. Risk is compositional too, depending on exposure, entitlement and what the asset is to the organisation, which means the answer draws on attributes that originate in different sources and have to be joined before anything can be sorted.

\item \textbf{Secrets and keys.} Who can read a secret, how many principals can, which secrets a compromised workload exposes. Read access is rarely granted directly. It arrives through role assumption, through resource policies, and through wildcards that subsume the specific grant, so counting readers means resolving entitlements to their effective population.

\item \textbf{Access and reachability.} The effective-permissions problem. Access is the union of identity-based and resource-based grants, modified by conditions and subsumption, and that union is materialised nowhere; it has to be computed. An answer that checks one mechanism is not partial but incorrect.

\item \textbf{Blast radius.} How far a compromise reaches, which is what makes something urgent. Reach is transitive, so the answer is a graph traversal, and error here is asymmetric: an underestimate de-prioritises exactly the asset that matters most.

\item \textbf{Control failures.} Which resources fail public-access controls broken down by control, which rank highest by number of failing controls, which fail encryption at rest. These are explicitly population questions. The answer is a count or a complete list, so sampling changes the answer instead of degrading it. Each control is also a predicate over configuration that several settings can satisfy or defeat.

\item \textbf{Exposure and cross-account.} The external attacker's view. Exposure is a property of a chain spanning network reachability, identity and data placement, and cross-account trust crosses the boundary of the account being inspected, so the relations that matter are the ones least likely to be visible from any single API.
\end{itemize}

\section{Evaluation Protocol}\label{sec:protocol}

All three arms answered the same 28 tasks against the same environment, read-only, and we scored final answers, not reasoning traces. Table~\ref{tab:config} states the configuration of each.

\begin{table}[ht]
\centering
\caption{Configuration of the three arms.}
\label{tab:config}
\small
\begin{tabular}{@{}l p{3.1cm} p{4.0cm} p{3.8cm}@{}}
\toprule
Arm & Model & Access & Run conditions \\
\midrule
Sola Security Brain & Claude Sonnet 4.6 & The context layer, queried & One process per task, five concurrent, warmed service, five-task warmup \\
\addlinespace
Claude Code & Claude Opus 4.8 & Read-only credentials and the provider command-line interface & \texttt{--max-turns} budget \\
\addlinespace
OpenAI Codex & GPT-6-Astra & Read-only credentials and the provider command-line interface & Claude Code's system prompt verbatim, 900s wall timeout \\
\bottomrule
\end{tabular}
\end{table}

The arms differ in model as well as in architecture, with the Sola arm on the cheaper Sonnet tier and both coding agents on their vendors' frontier models.

Coverage is measured by pooled relative recall over the claims the three systems make. Each step below is carried out by a language-model judge, and the whole chain is run three times under independent seeds \cite{Pradeep2025_GreatNuggetRecall}.

\begin{enumerate}
\item \emph{Extraction.} Each answer is reduced to the set of atomic factual claims it makes about the environment \cite{Min2023_FActScore,Wei2024_LongFormFactuality}.

\item \emph{Blind pooling.} The three claim sets for a task are merged into one joint pool \cite{SparckJones1975_IdealTestCollection,Voorhees2005_TRECBook}, labelled only SET-X, SET-Y and SET-Z under a per-task randomised permutation, so the judge cannot determine which system produced which set. A claim asserted by more than one system appears once in the pool but in each of those arms' claim sets, so the arms' claim sets intersect and the three coverages need not sum to one. Order is randomised against judge position bias \cite{Zheng2023_LLMJudge,Wang2024_NotFairEvaluators}.

\item \emph{Tier weighting.} Every claim in the pool is rated for security consequence on a four-level tier, L0 to L3, and weighted accordingly, with the rating made blind to which arm produced the claim. Claims the answers contradict each other on are neutralised where the disagreement is a matter of analysis rather than of configuration. The score therefore measures the share of security \emph{value} surfaced, not the number of claims counted \cite{Jarvelin2002_CumulatedGain,Nenkova2004_Pyramid}, and is graded rather than binary \cite{Lin2006_NuggetPyramids}.

\item \emph{Grounding.} A claim earns its weight in proportion to the faithfulness of the object it names, and only if that claim is actually contained in the arm's answer. Credit is the product of tier weight, grounded-object faithfulness and containment. This check runs on text scrubbed of arm fingerprints, not on relabelled sets alone \cite{Panickssery2024_SelfPreference}.
\end{enumerate}

For an arm $a \in \{\mathrm{sola},\mathrm{cc},\mathrm{cx}\}$, denoting the Sola Security Brain, Claude Code and Codex arms, with claim set $C_a$ over the joint pool $C = C_{\mathrm{sola}} \cup C_{\mathrm{cc}} \cup C_{\mathrm{cx}}$, with tier weight $w(c)$, grounded-object faithfulness $f_a(c) \in [0,1]$ and containment $g_a(c) \in \{0,1\}$, coverage is
\begin{equation}
\mathrm{Cov}(a) \;=\; \frac{\sum_{c \in C} w(c)\, f_a(c)\, g_a(c)}{\sum_{c \in C} w(c)} .
\end{equation}

The measure is \emph{relative recall} \cite{Clarke1997_RelativeRecall}: each system's share of a merged judged set, which is the standard construction where no enumerable gold set exists. It is not recall against the environment, and we do not report it as such. A claim that every system should have made, and none did, is invisible to it \cite{Zobel1998_HowReliable,Buckley2007_LimitsOfPooling}; we take this up in the limitations.
The four-step chain is itself model-driven, so every figure reported is the mean of three independent draws, each with its own blind assignment and seed, reported with the standard deviation across them. The ordering of the three arms was identical in every draw.

Alongside coverage we record two efficiency measures. Duration is wall-clock seconds per task, end to end. Cost prices each arm on output tokens at standard published rates. The Sola arm is priced on accumulated reasoning context at the Sonnet 4.6 rate of \$3/M, Claude Code on output tokens at the Opus 4.8 rate of \$15/M, and Codex on output tokens at the GPT-6 Astra rate. We additionally report metered cost for the two command-line arms, which prices input as well as output and applies the provider's long-context tier where a request exceeds it. The $5\times$ ratio between the Sonnet and Opus rates holds at both input and output tiers, so the pricing interpretation does not change the ratio between those two arms. The tokens priced are question-time inference tokens for all three arms.

Both measures are also reported normalised by the coverage they bought, as total seconds $\div$ total coverage and total dollars $\div$ total coverage. We normalise on suite totals rather than averaging per-task ratios, because a ratio with coverage in the denominator is unstable on tasks where an arm scores low.

\section{Results}\label{sec:results}

\begin{table}[ht]
\centering
\caption{Headline results.}
\label{tab:headline}
\begin{tabular}{llrrrl}
\toprule
\# & Measure & \shortstack[r]{Sola Security\\Brain} & \shortstack[r]{Claude\\Code} & \shortstack[r]{OpenAI\\Codex} & Result \\
\midrule
1 & Coverage (relative recall) & \textbf{0.549} & 0.340 & 0.281 & \textbf{$1.61\times$} CC, \textbf{$1.95\times$} Codex \\
2 & Duration, mean & 194s & 172s & 152s & slowest raw \\
3 & Duration per unit coverage & \textbf{353s} & 506s & 541s & \textbf{$1.4\times$ / $1.5\times$ faster} \\
4 & Cost per task & \textbf{\$0.0073} & \$0.1297 & \$0.1517 & \textbf{$17.8\times$ / $20.8\times$ cheaper} \\
5 & Cost per unit coverage & \textbf{\$0.013} & \$0.382 & \$0.540 & \textbf{$28.7\times$ / $40.6\times$ cheaper} \\
\bottomrule
\end{tabular}
\end{table}

\paragraph{Coverage.}

Headline results are given in Table~\ref{tab:headline}. Mean coverage is \textbf{0.549} against \textbf{0.340} for Claude Code and \textbf{0.281} for Codex, multiples of $1.61\times$ and $1.95\times$. Across three independent grading draws the figures vary by $\pm 0.012$, $\pm 0.011$ and $\pm 0.006$, and the ordering is identical in every draw. The Sola Security Brain leads 24 of the 28 tasks, Claude Code 3, and one is a tie between the Sola arm and Codex; Codex leads none outright. Because the pool is built from all three systems, the coverages share one denominator and are directly comparable.

\paragraph{Coverage by theme.}

\begin{table}[ht]
\centering
\caption{Coverage by theme.}
\label{tab:theme}
\begin{tabular}{lrrrrrr}
\toprule
Theme & $n$ & \shortstack[r]{Sola Security\\Brain} & \shortstack[r]{Claude\\Code} & \shortstack[r]{OpenAI\\Codex} & Gap vs CC & \shortstack[r]{Tasks\\led} \\
\midrule
Risk triage                & 7 & \textbf{0.527} & 0.367 & 0.304 & $+0.160$ ($+44\%$) & 7/7 \\
Secrets and keys           & 4 & \textbf{0.528} & 0.425 & 0.290 & $+0.102$ ($+24\%$) & 2/4 \\
Access and reachability    & 5 & \textbf{0.474} & 0.300 & 0.274 & $+0.174$ ($+58\%$) & 4/5 \\
Blast radius               & 4 & \textbf{0.600} & 0.278 & 0.288 & $+0.322$ ($+116\%$) & 4/4 \\
Control failures           & 4 & \textbf{0.563} & 0.345 & 0.245 & $+0.218$ ($+63\%$) & 3/4 \\
Exposure and cross-account & 4 & \textbf{0.638} & 0.313 & 0.270 & $+0.325$ ($+104\%$) & 4/4 \\
\bottomrule
\end{tabular}
\end{table}

Coverage by theme is given in Table~\ref{tab:theme}. The Sola Security Brain leads all six themes. The two coding agents sit closer to each other than either does to the Sola arm on five of the six themes, and on blast radius Codex overtakes Claude Code, 0.287 against 0.278.

\paragraph{Per-task coverage.}

Per-task results are given in Table~\ref{tab:pertask}, with the task text verbatim, sorted by the Sola Security Brain's advantage over Claude Code. Gaps against Claude Code range from $+0.58$ to $-0.15$. 7 tasks show a gap of 0.40 or more, 3 between 0.30 and 0.39, 15 between 0.01 and 0.29, and 3 are negative. The Sola Security Brain scores between 0.22 and 0.83, Claude Code between 0.10 and 0.59, and Codex between 0.08 and 0.54.

\begin{table}[ht]
\centering
\caption{Per-task coverage, sorted by advantage over Claude Code.}
\label{tab:pertask}
\small
\begin{tabular}{@{}l p{6.4cm} rrrr@{}}
\toprule
\# & Task & \shortstack[r]{Sola Security\\Brain} & \shortstack[r]{Claude\\Code} & \shortstack[r]{OpenAI\\Codex} & Gap \\
\midrule
q17 & Rank my resources by blast radius. & 0.74 & 0.16 & 0.11 & $+0.58$ \\
q20 & Which identities give the widest reach if compromised? & 0.81 & 0.34 & 0.40 & $+0.47$ \\
q12 & Which IAM roles can read my S3 buckets? & 0.69 & 0.25 & 0.21 & $+0.44$ \\
q23 & Which of my S3 buckets are publicly accessible? & 0.73 & 0.30 & 0.08 & $+0.43$ \\
q28 & Which resources are reachable from another account? & 0.68 & 0.25 & 0.11 & $+0.43$ \\
q03 & What are my riskiest compute assets? & 0.64 & 0.23 & 0.27 & $+0.41$ \\
q09 & Rank my secrets by how many principals can read them. & 0.66 & 0.25 & 0.18 & $+0.41$ \\
q27 & What are the attack paths to my privileged roles? & 0.83 & 0.44 & 0.46 & $+0.39$ \\
q25 & What are my most internet-exposed assets? & 0.61 & 0.23 & 0.28 & $+0.38$ \\
q14 & Which principals have wildcard access, and what can they reach? & 0.62 & 0.31 & 0.28 & $+0.31$ \\
q21 & Which resources fail public-access controls, broken down by control? & 0.62 & 0.36 & 0.15 & $+0.26$ \\
q24 & Which resources fail encryption-at-rest? & 0.61 & 0.35 & 0.42 & $+0.26$ \\
q05 & What should I fix first? & 0.50 & 0.27 & 0.26 & $+0.23$ \\
q02 & Who are my top 10 riskiest IAM identities? & 0.55 & 0.40 & 0.54 & $+0.15$ \\
q04 & What are my riskiest storage assets? & 0.44 & 0.29 & 0.25 & $+0.15$ \\
q06 & What are my highest-risk resource groups, and why? & 0.46 & 0.32 & 0.29 & $+0.14$ \\
q11 & Which internet-facing workloads can reach my secrets? & 0.53 & 0.39 & 0.23 & $+0.14$ \\
q13 & Who can decrypt my KMS keys? & 0.22 & 0.10 & 0.11 & $+0.12$ \\
q18 & If my most exposed resource is compromised, how far does it reach? & 0.42 & 0.30 & 0.38 & $+0.12$ \\
q19 & Walk me through my riskiest cluster. & 0.43 & 0.31 & 0.26 & $+0.12$ \\
q26 & Which exposed resources can reach sensitive data? & 0.43 & 0.33 & 0.23 & $+0.10$ \\
q07 & What are my riskiest secrets and encryption keys? & 0.62 & 0.59 & 0.34 & $+0.03$ \\
q16 & For my riskiest role, what can it reach? & 0.45 & 0.42 & 0.44 & $+0.03$ \\
q01 & What are my riskiest resources right now, and why? & 0.48 & 0.47 & 0.18 & $+0.01$ \\
q10 & If a workload is compromised, which secrets does it expose? & 0.49 & 0.48 & 0.49 & $+0.01$ \\
q15 & Which Lambda functions can be invoked, and by whom? & 0.39 & 0.42 & 0.33 & $-0.03$ \\
q22 & Rank my resources by number of failing controls. & 0.29 & 0.37 & 0.33 & $-0.08$ \\
q08 & Which secrets can more than one identity read? & 0.43 & 0.58 & 0.26 & $-0.15$ \\
\bottomrule
\end{tabular}
\end{table}

\paragraph{Duration.}

\begin{table}[ht]
\centering
\caption{Duration per task.}
\label{tab:duration}
\begin{tabular}{lrrr}
\toprule
 & \shortstack[r]{Sola Security\\Brain} & \shortstack[r]{Claude\\Code} & \shortstack[r]{OpenAI\\Codex} \\
\midrule
Mean                        & 194s & 172s & 152s \\
Suite total                 & 5{,}432s & 4{,}816s & 4{,}256s \\
\midrule
\textbf{Per unit of coverage} & \textbf{353s} & 506s & 541s \\
\bottomrule
\end{tabular}
\end{table}

Durations are given in Table~\ref{tab:duration}. On raw wall-clock the Sola Security Brain is the slowest arm: 194s per task against 172s for Claude Code and 152s for Codex.
Raw wall-clock prices different quantities of work at the same rate. Normalised by the coverage it bought, the ordering reverses completely. The Sola Security Brain spent 5{,}432 seconds to deliver 15.37 units of coverage, or \textbf{353s per unit}, against Claude Code's 4{,}816 seconds for 9.51 units (\textbf{506s}) and Codex's 4{,}256 seconds for 7.87 units (\textbf{541s}). Codex is the fastest arm per task and the slowest per unit of coverage. A system that answers a population task by enumerating the population spends longer than one that samples and stops, and the coverage column is where that time reappears.
The Sola arm completed every task: 28 of 28 finished, with no empty answers and no format or validation retries.

\paragraph{Cost and coverage-normalised efficiency.}

\begin{table}[ht]
\centering
\caption{Reasoning cost and efficiency.}
\label{tab:cost}
\begin{tabular}{lrrr}
\toprule
 & \shortstack[r]{Sola Security\\Brain} & \shortstack[r]{Claude\\Code} & \shortstack[r]{OpenAI\\Codex} \\
\midrule
Mean output tokens per task      & \textbf{2{,}448} & 8{,}646 & 3{,}033 \\
Mean input tokens per task       & not instrumented & not instrumented & 314{,}311 \\
Normalised cost per task         & \textbf{\$0.0073} & \$0.1297 & \$0.1517 \\
Normalised cost, 28 tasks        & \$0.204 & \$3.632 & \$4.248 \\
\midrule
\$ per unit of coverage          & \textbf{\$0.013} & \$0.382 & \$0.540 \\
Metered cost per task            & --- & \$0.5849 & \$1.0936 \\
\bottomrule
\end{tabular}
\end{table}

Costs are given in Table~\ref{tab:cost}. The ratio against Claude Code decomposes cleanly: $3.5\times$ fewer output tokens, consistent with a context-graph loop reasoning over compact retrieved facts rather than a growing shell transcript, multiplied by the $5\times$ lower model rate. Codex inverts the intuition its output count invites. It emits 3{,}033 tokens per task, close to the Sola arm's 2{,}448, but an agentic command-line loop re-feeds the provider's output on every turn: Codex consumed a mean of 314{,}311 input tokens per task, a ratio of roughly 104 to 1, and twelve of the 28 tasks crossed the long-context threshold and billed the whole request at the higher tier. On metered rates Codex is the most expensive arm in the study, \$1.0936 per task against Claude Code's \$0.5849. The normalised figures are the conservative like-for-like comparison and are what we report; the Sola arm's input tokens were not instrumented, so no metered figure is claimed for it.

\section{Discussion}\label{sec:discussion}

The widest margins cluster on one task shape: \emph{which}, \emph{how many}, \emph{rank these}. Blast-radius ranking, widest identity reach, which roles can read storage, what is reachable from another account, secrets by reader count. These are properties of a population, not of a resource, and answering them means enumerating the complete set: unioning identity-based grants with resource policies, applying wildcard and administrative subsumption, resolving grants to their real target populations, sweeping every exposure family. Both coding agents recompute fragments of that live from raw APIs and exhaust their budget on breadth; in the answers that disclose a sample, roughly forty resources of a type were inspected. Codex samples more shallowly still and stops earlier. Its weakest themes are control failures and exposure with cross-account reach, at 0.245 and 0.270, both of which are answered by sweeping a complete set rather than by inspecting a sample.

One pattern recurs in both agents' answers, and we call it \textbf{sample-and-generalise}. First, the agent inspects a fraction of a large population. A bounded budget makes this rational, and choosing what to look at is the task, so this part is not a defect. Second, it reports on the whole population rather than on the fraction: the answer states that a property holds of nothing in the environment, not that it held of nothing in the subset examined. The claim is universal, the evidence is partial, and nothing in the sentence marks the difference. Third, the sample size is disclosed, but in a scope-caveats field below the answer rather than in the sentence making the claim. Nothing is concealed, and a reader who looks will find it, but a caveat filed below a conclusion is not weighed against it.

Together these produce a failure the reader cannot catch. A fabricated fact can be checked against the environment and found false. A finding that was never surfaced leaves nothing to check: a universal negative reads identically whether the agent enumerated the population or a fraction of it, so a true negative and a false one arrive as the same sentence. We describe this pattern, we do not measure it: our tasks were not stratified by population size, and we report no incidence rate.

Task q12, \emph{which IAM roles can read my S3 buckets}, is the clearest instance. Claude Code stated that no bucket policies exist, having checked four bucket families that all returned \texttt{NoSuchBucketPolicy}, and concluded there were no public or cross-account grants. The scope-caveats field of the same answer records that 40 of roughly 5{,}000 buckets were sampled. The answer is internally honest and externally wrong.

S3 read access is granted through several mechanisms, so an answer that checks bucket policies alone misses most of the grants. The Sola arm answered in three tiers: the administrative and power-user tier, which subsumes all S3 read by construction; the managed read-only tier, across 189 matching rows; and the specific grants, an inline Lambda-role family across 1{,}410 matched rows plus the distinct principals named in bucket policies. Sixty-five buckets carry a \texttt{Principal: *} read grant. Five external accounts hold cross-account read, one of them across 53 buckets. Whether any given wildcard grant is reachable depends on block-public-access state and policy conditions, which is triage an analyst can perform. It is not triage they can perform on a category reported as absent.

Claude Code leads three tasks, and a fourth is a tie that it loses. It takes q08, \emph{which secrets can more than one identity read}, at 0.58 against 0.43; q15, \emph{which Lambda functions can be invoked, and by whom}, at 0.42 against 0.39; and q22, \emph{rank my resources by number of failing controls}, at 0.37 against 0.29. On q10, \emph{if a workload is compromised, which secrets does it expose}, the Sola arm and Codex tie at 0.49. All four sit in reach the layer does not pre-index: the set of secrets carrying more than one reader, the path from a workload to the secrets it can read, the invocation surface of a function, and a count-based ranking with no precomputed aggregate. Where there is no materialised relationship to sweep, the answer depends on which resources the live agent happens to sample, and the advantage tracks what has been modelled rather than the architecture in the abstract \cite{Xiang2025_WhenGraphsRAG}. That is a coverage boundary rather than a ceiling: the missing edges can be materialised.

Everything above was measured on a single cloud vendor, which is the setting most favourable to the coding agents. AWS is the provider a context-free model knows best, its API surface is uniform, and one set of credentials reaches the whole estate, so the agent can reconstruct structure from a single coherent schema. Real environments are not one vendor. Identity spans a directory, a cloud and a set of SaaS applications; a workload's data reaches a warehouse governed elsewhere; the same principal appears under different identifiers in each system, and no API joins them. Reconciling those entities and inferring the relations between them is the work the modelling layer does offline, and it is work a live agent must repeat per question against schemas it was not designed around. We did not test this, so we state it as an expectation rather than a result: the gap we report should be read as a lower bound, and we would expect it to widen on multi-vendor estates rather than narrow.

That is the case for the layer, and it is an argument about when the work is done rather than about how well any of these systems reason. Completeness cannot be added at inference time; the representation has to hold the population before the question arrives. Normalisation, entity reconciliation and relation inference are done once and amortised across everything later asked of the layer, where a live agent pays that cost on every task and pays it again on the next. The Sola arm carried 2{,}448 output tokens per task against Claude Code's 8{,}646, a factor of 3.5; the remaining factor in the $17.8\times$ cost ratio follows from the model tiers the arms run on rather than from the architecture. Codex makes the point from the other direction. It is the most expensive arm on metered rates and returns the least coverage, because reconstructing structure live means re-feeding the provider's output on every turn, and that traffic is billed whether or not it advances the answer. The layer does not win by spending more; on every basis available to us it spends least.

Several constraints bound what this evaluation supports. Coverage is relative recall over a pool built from the three systems' own output rather than against an external oracle, so it cannot distinguish one arm finding more of what matters from all three missing the same things, and the denominator is defined by the systems compared rather than by the environment \cite{Zobel1998_HowReliable,Buckley2004_IncompleteInformation,Buckley2007_LimitsOfPooling}. A claim that every system should have made, and none did, is invisible to the measure. Three arms make the pool less dependent on any one pairing than two did, and the ordering held in every grading draw, but the pool is still built from systems rather than from the environment. Only a reference set drawn independently would close that.

The clearest next step is more coding agents. Second, ablation. Holding the task set fixed and varying one thing at a time would attribute the gap to its sources: the model behind each arm, the components of the security brain, and the customisations available to a coding agent. The last matters most for fairness, since a coding agent is not fixed at its default. Skills supply domain procedure on demand \cite{Anthropic_AgentSkills}, and MCP servers connect an agent to systems beyond one vendor's CLI \cite{MCP_Spec}. An agent equipped with security skills and a set of vendor MCP servers is a stronger baseline than either of the ones we ran, and the honest version of this comparison tests against it. Third, an independent reference set: an assessor, or an independent pass over the environment, would propose the claims a complete answer should contain, and coverage reported against that fixed set would convert our principal methodological threat into a measured quantity. Fourth, the graph-disabled ablation, which would isolate the layer's contribution within a single arm.

\section{Conclusion}\label{sec:conclusion}

Over 28 cloud-security investigation tasks on one live AWS environment, an enterprise security brain reached coverage of $0.549 \pm 0.012$ against $0.340 \pm 0.011$ for Claude Code and $0.281 \pm 0.006$ for OpenAI Codex, scored on a single pool built from all three, with the ordering identical in every grading draw. It led 24 of the 28 tasks, from the cheapest model tier and at an eighteenth of Claude Code's cost per task and a twenty-first of Codex's. On raw wall-clock it is the slowest of the three; per unit of coverage it is faster than both and cheaper by $28.7\times$ and $40.6\times$.

Both coding agents reason capably over cloud infrastructure and report their own limits honestly. Their coverage of population tasks is nonetheless bounded by what they can enumerate within a budget, and neither a frontier model on either side nor a second vendor's agent lifted that bound. Beyond the aggregate, we described an answer-level pattern we call \emph{sample-and-generalise}, seen in both agents: partial enumeration, an unhedged universal negative, and a sample size disclosed only in metadata. We did not measure how often it occurs. A false negative and a true negative are the same sentence, so the failure is not detectable from the claim itself, and the disclosure that would expose it sits where a reader will not weigh it. That is what makes it consequential in a domain where the finding that was not surfaced is the one that gets exploited.

What would strengthen the claim further is a wider basis still: more coding agents in the pool, an ablation that varies the model, the layer's components and the agent's own customisations one at a time, and a reference set of claims drawn independently of the systems under test.
\bibliographystyle{unsrtnat}
\bibliography{references}

\end{document}